# Ultrafast voltage-tunable detectors for Terahertz radiation operating above 100K

G. B. Serapiglia[1], M. Hanson[2], M. F. Doty[1*], P. Focardi[3], W. R. McGrath[3], A. C. Gossard[2] and M. S. Sherwin[1]

[1]*Physics Department and iQUEST, University of California, Santa Barbara, CA 93106*

[2]*Materials Department, University of California, Santa Barbara, CA 93106*

[3]*Submillimeter-Wave Superconductive Sensors Group, Jet Propulsion Laboratory, Pasadena, CA 91109*

* Present address: US Naval Research Lab 4555 Overlook Ave., SW Washington, DC 20375

Collective vibrations of proteins[1], rotations of small molecules[2], excitations of high-temperature superconductors[3], and electronic transitions in semiconductor nanostructures[4] occur with characteristic frequencies between 1 and 10 THz. Applications to medicine[5], communications[6], security[7] and other fields are emerging. However, mapping the coldest parts of the universe[8] has been the largest driver for developing THz detectors. The result is a family of exquisitely-sensitive detectors requiring sub-4K temperatures[9]. For earthbound THz science[10] and technology, sensitivity remains important but many applications require high speed and operating temperatures. Room-temperature Schottky diodes enable some of these applications[11]. Here we demonstrate a new type of detector in which THz radiation excites a collective oscillation of ~25,000 electrons between two gates in a microscopic four terminal transistor. The energy dissipates into other modes of the electron gas, warming it and changing the source-drain resistance. The detector shows amplifier-limited rise times near 1 ns and has detected THz laser radiation at temperatures up to 120K. The frequency of the collective **oscillation tunes with small gate voltages. The first-generation tunable antenna-coupled intersubband Terahertz (TACIT) detectors[12] tune between 1.5 and 2 THz with voltages <2V.**

The power of band-gap engineering in semiconductors has not made a large impact on the problem of THz detection. In contrast, quantum well infrared photoconductors (QWIPs)[13] are very successful in the 25-38 THz (8-12 μm) frequency range. Handheld video cameras with large-format focal-plane arrays of QWIPs cooled to 70K by mechanical coolers are commercially available. Very recently, QWIPs and related devices[14] have been demonstrated with response between 3.5 THz[15] and 7 THz[16]. These operate best below 20K. There exist semiconducting detectors which are frequency-tunable with magnetic field, including one operating at temperatures below 0.1 K that has detected single THz photons[17]. Macroscopic field-effect transistors containing 2-D electron gases (2DEGs) in Si MOSFETs,[18] at a AlGaAs/GaAs heterojunction[19] and in asymmetric coupled quantum wells[20,21] can be voltage-tunable detectors operating between 1 and 5 THz. However, they have not become competitive devices, largely because of relatively poor coupling efficiency of THz radiation into the 2-DEG.

The TACIT detector reported here [12] enhances coupling by monolithically integrating a receiving antenna with a microscopic electronic absorber consisting of a 2-DEG in a GaAs/AlGaAs quantum well with a resonant intersubband absorption at THz frequencies. The antenna is capacitively coupled to the 2-DEG, eliminating series resistance losses associated with ohmic coupling in transmitting the power between the antenna and load. With 4 terminals, the device's resonant frequency can be tuned over a broad band while maintaining independent control of the charge density $N_S$ in the 2-DEG. The impedance that the active region presents to the antenna depends on $N_S$ and other parameters, and can in realistic future designs be made to match the impedance of the antenna[22].

Fig 1a shows schematically a device glued at the focus of a silicon lens. The lens couples the impinging THz beam to the antenna radiation pattern of a twin slot antenna[23], shown in fig 1b. The electromagnetic fields received by each slot antenna propagate on a coplanar wave-guide.  One of these terminates on a back gate, the other on a front gate (fig 1c). In between these 4.5x4.5 $\mu$m gates is a 0.7 $\mu$m thick section of a dumbbell-shaped "mesa" composed of an $Al_{0.7}Ga_{0.3}As$ barrier, a pair of coupled, remotely-doped GaAs quantum wells and another $Al_{0.7}Ga_{0.3}As$ barrier. The THz electric field generated across the quantum wells is in the growth direction, and excites the intersubband plasmon resonance where charge oscillates back and forth between the two coupled quantum wells. The intersubband excitation dissipates energy into in-plane modes of the 2DEG via interface and impurity scattering, and to the lattice via optical phonon scattering on timescales between 1 and 1000ps, depending on temperature[20,24]. In this way radiative energy transfers to the 2DEG and heats it, increasing the monitored resistance to currents flowing between the source and drain.

 Three TACIT devices from the same manufacturing batch were tested using three sources of THz radiation with different time structures.  The sources were a molecular gas laser with a continuous beam, UCSB's Free-Electron Lasers (FEL) with 5 $\mu$s pulses, and "slices" of the FEL output[25]  with ~ns duration.  FEL pulses were typically attenuated by 1000 or more.  The fast response of a TACIT sensor to THz pulses with durations 1.5 and 3.5 ns is shown in Fig. 2.

Frequency-dependent intersubband absorption data on macroscopic devices measured at T=1.5K showed that the intersubband absorption was voltage-tunable from 1.2 to >2 THz, and exhibited a full-width at half max (FWHM) of 300 GHz at 1.53

THz. The resonant nature and voltage-tunability of the TACIT sensors' response to THz radiation are demonstrated in Figs. 3a and c. A THz beam of fixed frequency illuminated the sensor as the dc electric field $E_{DC}$ in the growth direction was varied. Front- and back-gate voltages were applied to the center conductors of the CPW lines that terminate on the active region to generate $E_{DC}$=( [$V_{Front\ gate}$-$V_{back\ gate}$]/gate separation. In all of the experiments, the front gate, back gate, and source-drain voltages were adjusted to vary $E_{DC}$ while maintaining a constant charge density in the active region. All curves show two clear maxima in the photoresponse, roughly symmetric about $E_{DC}$=0.

The two maxima in the photoresponse occur close to the two values of $E_{DC}$ at which the intersubband absorption is calculated to be resonant with the THz laser frequency (arrows in Fig. 3b,c). Fig. 3b shows the intersubband absorption frequency as a function of $E_{DC}$ calculated using a self-consistent Poisson-Schrödinger solver, taking into account both the temperature and the electron-electron interaction in the Hartree or random phase approximation[26]. The near-symmetry of the calculated intersubband absorption frequency about $E_{DC}$=0 is explained in Fig. 3d. In the left (right) figure the (left) right well is higher in energy, but the spacing between the two energy levels is nearly the same in both cases. The minimum intersubband absorption frequency is determined by the tunneling rate from one well to the other through the thin $Al_{0.3}Ga_{0.7}As$ barrier.

The positions of experimental and theoretical peaks in photoresponse match calculations. The solid lines in Fig. 3c are Lorentzian functions $A*\Gamma/[(f\ -f_0)^2 + \Gamma^2]$ calculated at each field, where A is a scale factor to match the theoretical and experimental maxima, $f_0$ is the resonant frequency at each field predicted by the tuning curve in fig 3b and $f_L$ is the fixed laser frequency. The half-width-at-half-maximum $\Gamma$ is estimated by adding a 110 GHz contribution from electron-phonon scattering at 70 K [Ref. 27], to a field dependent contribution of 2.5 THz/(mVAngstrom[-1]) . The latter is largely due to interface scattering, and is estimated from FTIR data[24]. The experimental peaks are wider than the model predicts, and there may be a nonresonant background. The explanation of these discrepancies is beyond the scope of this Letter.

Appropriate figures of merit for a direct detector include responsivity R (V/W) and noise-equivalent power (NEP, W/Hz$^{1/2}$), the power needed for detection with a signal-to-noise ratio of 1 in a 1 Hz bandwidth. The minimum detectable power in a 5 μs pulse, correcting only for the transmission of windows and Si/air interface, at a bath temperature $T_b$=80K, was 1 mW using 1 MHz bandwidth amplifier, corresponding to noise-equivalent power (NEP) of 10$^{-6}$ W/Hz$^{1/2}$. Noise in this measurement was dominated by sources extrinsic to the detector.

The internal responsivity of the TACIT detector is defined as voltage per unit power dissipated in the active region. This internal responsivity is estimated in three ways: from the optical responsivity $\boldsymbol{R}_{optical}$(V/ incident THz power), accounting for large estimated losses; from the nonlinear source-drain I-V curve; and from a simple theory. The two estimates based on experiment agree within experimental error, while the theoretical estimate is about one order of magnitude higher. Discrepancy between experiment and theory will be investigated in future studies.

The measurement of $\boldsymbol{R}_{optical}$ for sample A begins with a heavily-attenuated beam of 5 μs pulses from the UCSB FEL with a power of 10's of mW. Dividing the induced source-drain voltage by the incident power (outside the cryostat), we find R$_{optical}$=(30±15) mV/W. Severe losses, which can be mitigated in future devices, attenuate the incident optical power on its way to the active region. The cryostat windows transmit 40±1% of the beam, and the interface between air and the Si lens transmits 70%. The beam does not match the antenna pattern of the twin slot antenna, leading to a coupling efficiency of (4±2)% into the antenna mode. The antenna has a source impedance $Z_{Source}$=20Ω-i40Ω, (see supplementary materials). The load impedance[16] presented by the active region is $Z_{Load}$~(1.7±1.2)Ω-i(31±9)Ω , calculated for the intersubband absorption tuned to 1.53 THz and electron temperature 120K. Largely a result of the capacitive source and load impedances, only

$1-\left|\dfrac{Z_S-Z_L^*}{Z_S+Z_L}\right|^2$ =(2.5±0.3)% of power is transferred from the antenna to the load. The

product of these losses is $\eta$=(2.8±1.4)x10⁻⁴, where errors are assumed independent and added in quadrature. Dividing the optical responsivity by the product of the losses, we deduce an internal responsivity of $\boldsymbol{R}_{optical}/\eta$= 107±75 V/W at a bath temperature of 80 K. Additional losses (for example, associated with not placing the TACIT detector at the exact center of the Si lens), cannot be excluded.

The internal responsivity can be estimated experimentally from the I-V curve. For conventional bolometers, this is called the electrical responsivity $\boldsymbol{R}_{electrical}$. The simplest expression for this quantity is $\boldsymbol{R}_{electrical}$ =(V/I-dV/dI)/2V[9]. Fig. 4 shows the source-drain I-V curve and the differential resistance derived from differentiating a fit to the I-V curve. At the operating point, R$_{eletrical}$=400±150 V/W.

The experimental values for the internal responsivity can be compared to a theoretical calculation based on a standard bolometric model[12]

$$\frac{dV}{dP} = I \frac{dR}{dT_e}\frac{dT_e}{dP} = \frac{V_o\gamma\tau}{C_v} \approx \frac{V_o\gamma\tau}{N_S A k_B}.$$

(1)

Here $dV/dP$ is the rate of change of voltage with incident power, $I$ is the bias current, $V_0$=0.07 V is the bias voltage, $(1/R)dR/dT_e$ =$\gamma$ is the normalized rate of change of device resistance with electron temperature $T_e$, $\tau$ is the electron energy relaxation time, $C_V$ is the heat capacity of the electrons in the active region, $N_S$ =1.3±0.5 x10¹⁵ m⁻² the areal density of the carriers, $A$ =(4.5 μm)$^2$ is the area, and $k_B$ is Boltzmann's constant. The final denominator is the heat capacity in the classical 2D approximation. Inset to Fig. 4 is a plot of the device resistance as a function of temperature, showing that it increases at a maximum rate above 60 K when optical phonon scattering is energetically activated. Based on comparing the differential resistance at the operating point with the temperature-dependent low-current resistance, we deduce $T_e$=120K and $\gamma \tilde{} 1$.Note that, in samples with higher mobility, $\gamma$ can be significantly enhanced (blue curves in inset to Fig. 4) Assuming bulk LO phonon scattering is dominant, we deduce from Seeger[28] (Eq. 6.12.20) an electron energy relaxation time of 0.6 ps. This model and set of parameters predict a responsivity $\boldsymbol{R_{theoretical}}$=(1100±400) V/W from $N_S x A$=25,000 electrons in the active region.

This Letter has demonstrated the feasibility of TACIT sensors, and points the way to improving their performance by many orders of magnitude (see supplementary materials). With improvements in coupling and to the active region, NEP < 1 pW/Hz$^{1/2}$ can be anticipated with <10 ps rise times at $T_b$=50K. Such a NEP is sufficiently low to directly detect weak THz emissions from room-temperature objects with high signal to noise ratio. This temperature is accessible to liquid-free cryocoolers (see Sumitomo SRS 2110) which consume <60W, weigh about 2 kg, and have ~10 cm characteristic linear dimensions. With their unprecedented combination of portability, sensitivity, speed and tunability, future TACIT detectors can enable new time-resolved experiments; light, compact, nearly quantum-limited heterodyne receivers with ultra-wide intermediate frequency bandwidths[22]; focal plane arrays for passive and active imaging; spectrometers on a chip; THz communications; and THz radar.

**Supplementary Information** accompanies the paper on *Nature*'s website (http://www.nature.com).

Acknowledgments:  This work was supported by NASA and the NSF.  We thank B. Thibeault for assistance with process development, S. Carter for assistance with data acquisition programming, Yongqing Li for scanning electron micrographs, and D. Allen, E. McFarland, K. Plaxco, J. Rothman and C. Weinberger for critical readings of the manuscript.

Author contributions:  During 4 years as a post-doctoral researcher, GBS developed the process to fabricate TACIT sensors, and performed all experimental measurements.  MH grew the GaAs/AlGaAs samples from which TACIT sensors were fabricated.  MFD developed and operated the THz pulse slicer and acquired data of Fig. 2 with GBS.  PF performed calculations of the antenna source impedance and efficiency of THz beam coupling into beam pattern of TACIT detector.  WRM, in collaboration with MSS and GBS, designed the twin slot antenna and embedding circuit for the TACIT detector.  ACG is MH's Ph. D. supervisor and MSS's long-time collaborator for MBE-grown samples.  MSS conceived of and developed the theory of the TACIT detector, and co-ordinated and closely supervised the entire project.

Competing interests:  MSS owns U. S. Patent 5,914,497 for the TACIT detector.

Correspondence should be addressed to MSS. (e-mail: sherwin@physics.ucsb.edu.).

Fig. 1:  Schematic diagrams and scanning electron micrographs (SEM) of TACIT detectors.  A)  A THz beam is focused by converging lens onto the TACIT detector.  Actual Si lenses used were 12 mm or 6mm in diameter.  B) Scanning electron micrograph of a TACIT detector, showing H-shaped "twin slot antenna" designed for receiving 1.6+/- 0.4 THz.  THz waves are prevented from flowing away from the active region by rf choke filters.  These consist of a series of 1/4-wavelength-long sections of center conductor which alternate between two different widths (and hence impedances). C)  Schematic diagram and inset SEM showing active region of TACIT detector.  The dumbbell-shaped "mesa" is mostly made of $Al_{0.3}Ga_{0.7}As$ (green) except for the blue and red regions representing cold and warm sections of the 2-D electron gas (2DEG) which is trapped in the GaAs quantum wells.  Alloy spikes from the THz ground plane make Ohmic contact to the cold 2DEG, forming 100µm x 100µm source and drain.  The inset to Fig. C shows a SEM close-up of the front gate.  See supplementary materials for how devices were fabricated.

Fig. 2:  Response of a TACIT detector (device A) at bath temperature $T_B$=77K to pulses with frequency 1.53 THz, power ~100mW  and duration 1.5 ns (blue curve) and 3.5 ns (red curve).  These pulses were generated using the UCSB Free-Electron Laser and "pulse slicer."  The ~1 ns rise time is limited by electronics. Measurements and theories of the intersubband relaxation time imply that actual rise and fall times are below 10 ps at 77K.[20,29] A signal has been detected at lattice temperatures as high as 120K using the 5 µs FEL pulses.  The electrical noise is not intrinsic, but was radiated from one of the lasers used for the pulse slicer.  The reason for the elevated baseline after the pulses is not clear.  The source-drain bias and source-drain resistance were 0.12 V and 522 Ohms, respectively.  The displayed voltage is the output of a preamp with a gain of 100 and input impedance 50 Ohms.

Fig. 3: Voltage-tuning of the TACIT detector. A) Photosignal vs. dc electric field in response to 148 Hz chopped CW radiation from a molecular gas laser, average power ~3mW and tuned to 1.63 THz for device A at $T_B$=77K (blue curve); and 25mW peak power, 5 µs pulses from the UCSB Free-Electron Laser tuned to 1.53 THz for device B at $T_B$=80K (red curve). The source-drain bias was ~0.1 V for device A and ~0.07V for device B, and the quoted laser powers have not been corrected for any losses. Sample A was mounted on a 6mm diameter lens, and B on a 12 mm lens, with THz incident as shown schematically in Fig. 1a. B) Intersubband resonance frequency vs. dc electric field calculated for $T_e$=60 K with charge density $8 \times 10^{10}$ cm$^{-2}$. Horizontal lines are FEL frequencies 1.53 and 2.0 THz used to measure the photoresponse of sample C. C) Photoresponse of sample C to 5 µs pulses of 1.6 (blue squares) and 2.0 (red squares) THz radiation from the UCSB FEL, with $V$=0.12 V, $T_b$=15K, and $T_e$~68K. The solid lines show the predicted photoresponse based on the convolution of the tuning curve in B) with a Lorentzian with FWHM=0.5 THz. D) Conduction band energy vs. position for the quantum well in the active region (black), and wave functions of the lowest-energy states of the two lowest subbands ( blue and red) for dc electric fields -20 kV/cm (left) and 20 kV/cm (right).

Fig. 4: Voltage vs. source-drain current for sample A at $T_b$=80K (black dots), 9[th] order polynomial fit to V-I (black line), and dV/dI from differentiating fit. Inset: Source-drain resistance as a function of temperature for a TACIT device (red curve) and for a higher mobility sample (solid black curve). The solid curve was calculated by scaling our sample data by the mobility measured in Ref. [30]. Also shown is *1/R(dR/dT)* for our sample and for the higher-mobility sample.

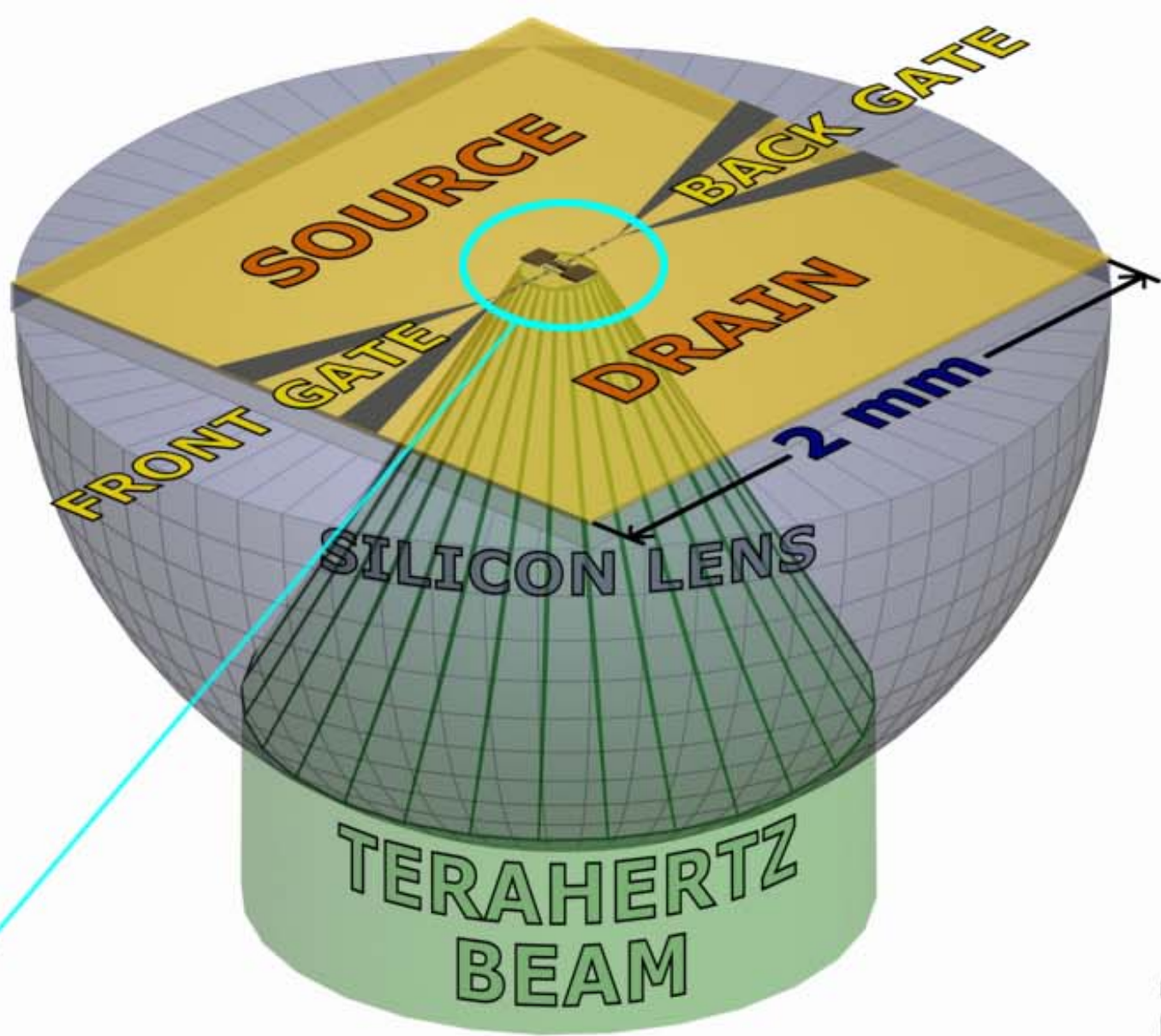

SOURCE
BACK GATE
FRONT GATE
DRAIN
2 mm
SILICON LENS
TERAHERTZ
BEAM

**a.**

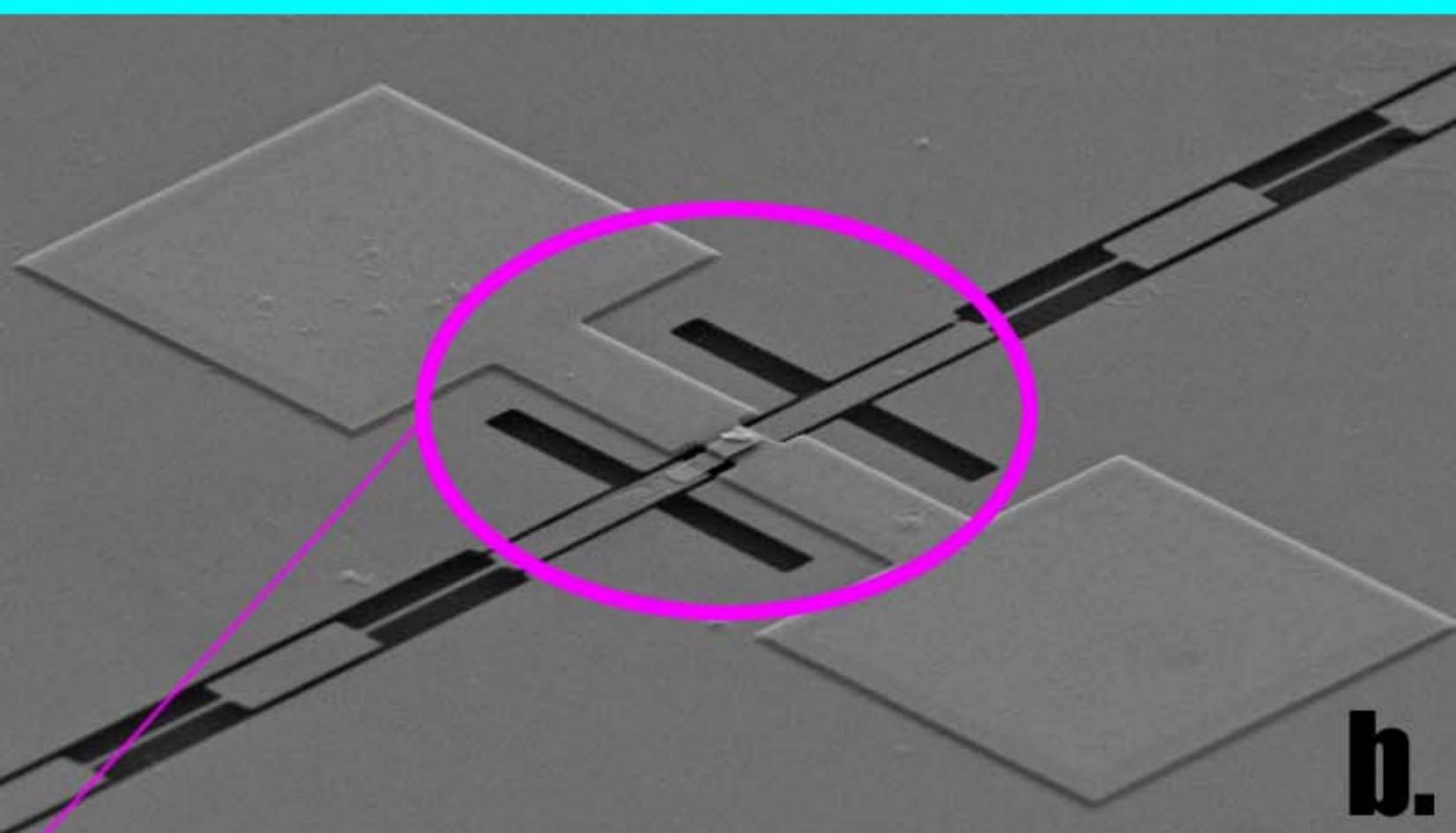

**b.**

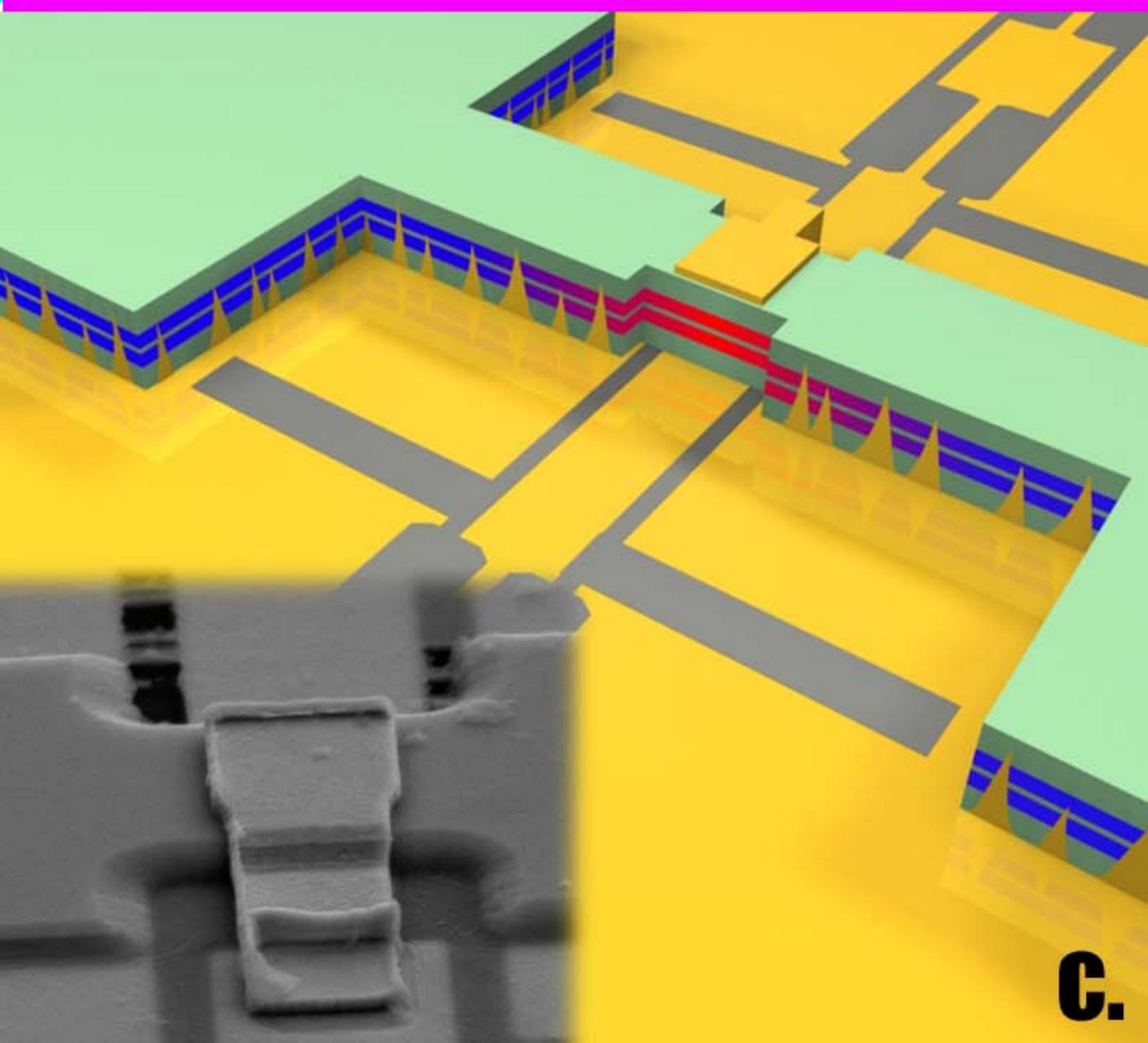

**c.**

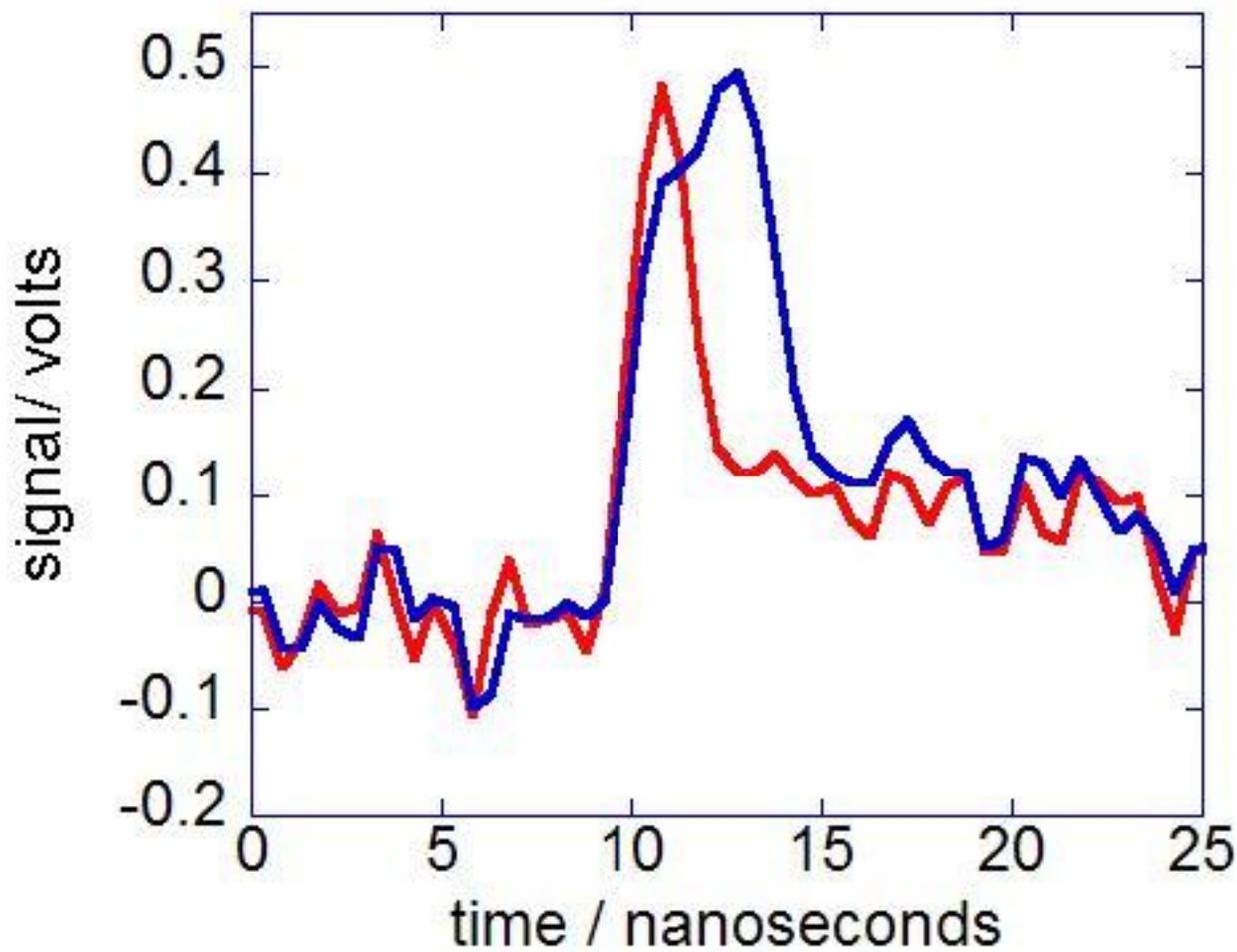

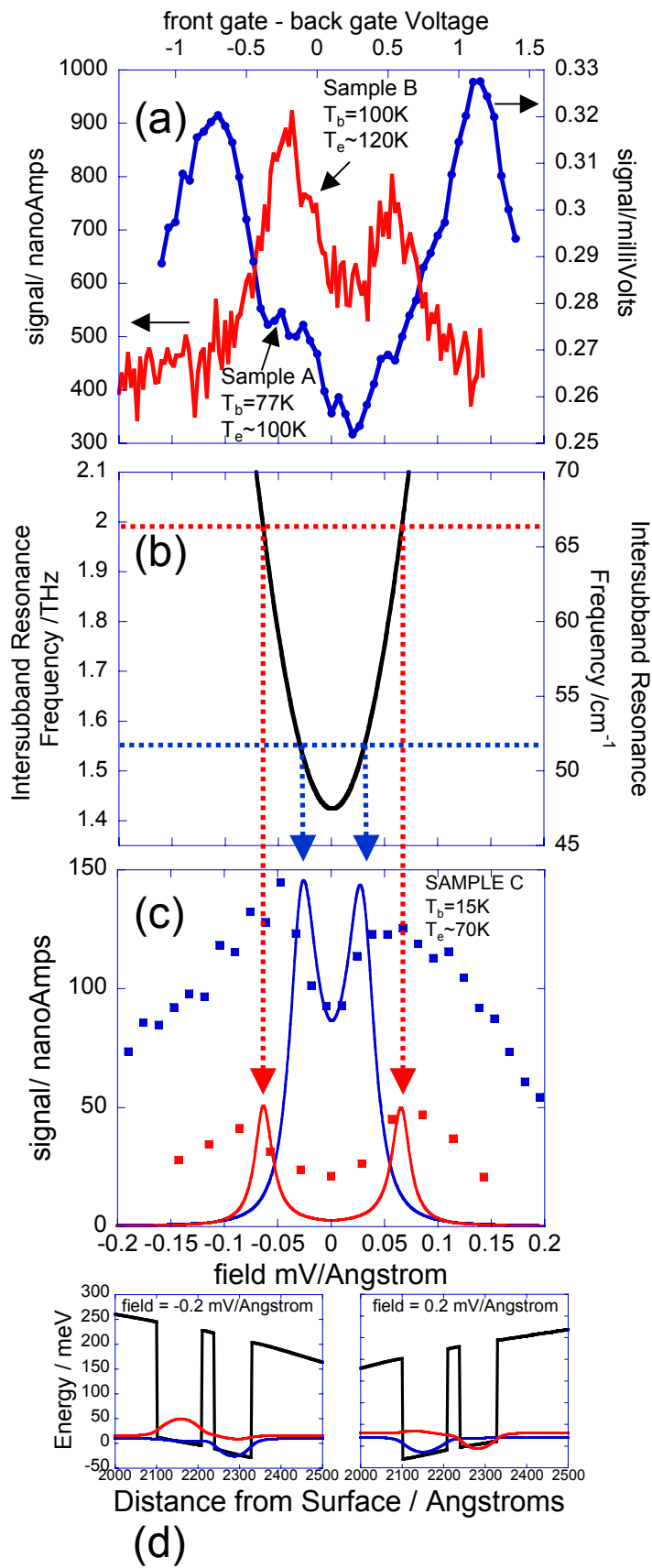

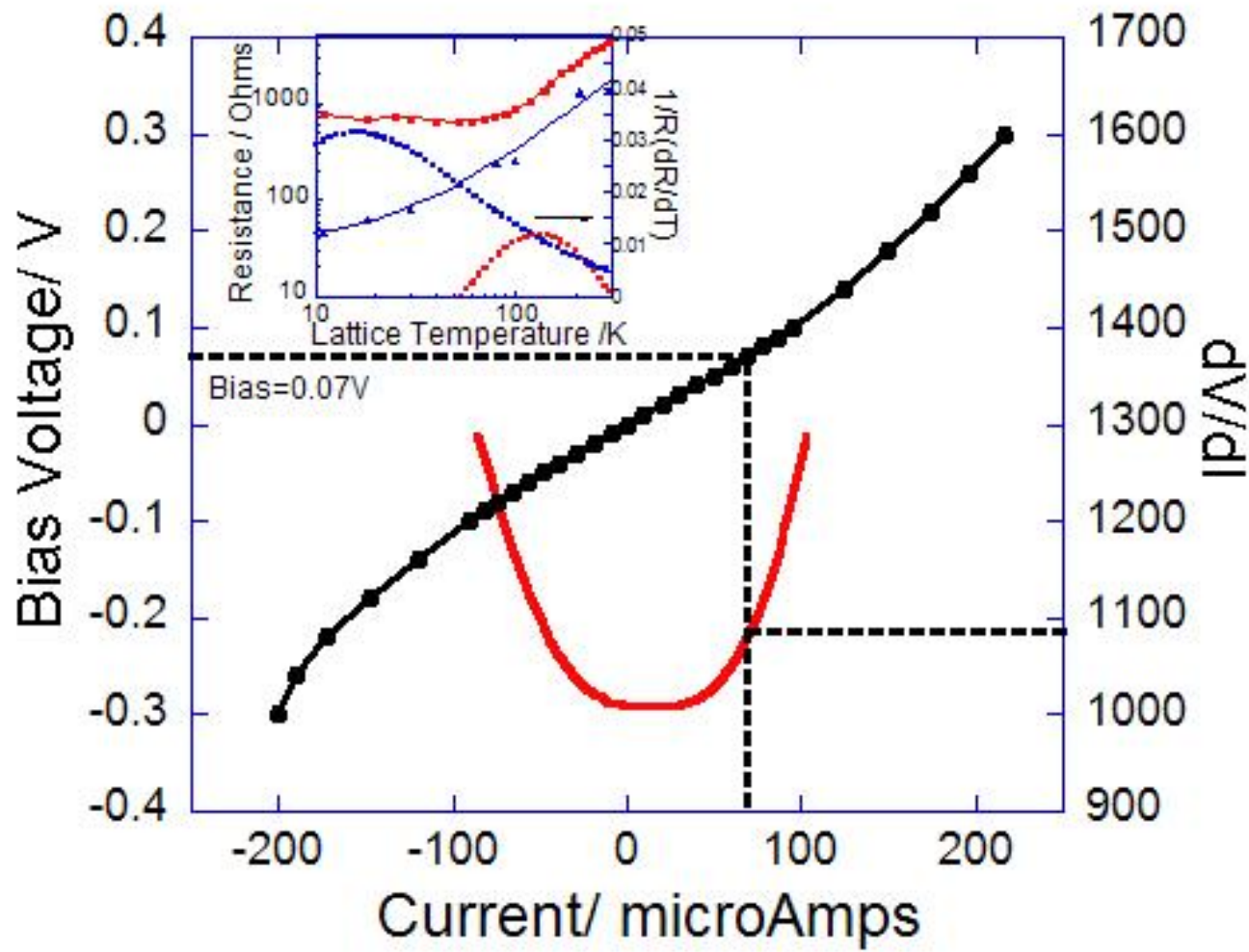

**Supplementary material**

*Sample growth and device fabrication*

The TACIT detector active region consists of the following heterostructure grown by Molecular Beam Epitaxy: starting at the surface, a 10nm GaAs cap layer, 130nm $Al_{0.3}Ga_{0.7}As$, 1 monolayer Si delta doping, 70nm $Al_{0.7}Ga_{0.3}As$, a 10.8nm GaAs well, a 3nm $Al_{0.3}Ga_{0.7}As$ barrier, a 9nm GaAs well, 70 nm $Al_{0.3}Ga_{0.7}As$, 1 monolayer Si delta doping, 197nm $Al_{0.7}Ga_{0.3}As$, 200nm GaAs and a 1 micron $Al_{0.7}Ga_{0.3}As$ etch stop layer. In order to fabricate our device, we used standard stepper photo-lithography. An outline of the process, based on the epoxy-bond and stop-etch (EBASE) technique[1] is as follows: First two Au/Ge/Ni/Au 100 x 100 μm ohmic contact pads were defined, metallized and annealed, separated by 5μm, for the source and drain. Next, the twin-slot antenna, co-planar waveguides, filters and Schottky gates were deposited in one Ti/Au metallization step. The wafer was subsequently glued processed side down on a host GaAs wafer. In this way it was possible to chemically remove the 0.5 mm thick substrate of the processed wafer with one selective etch stopping at the $Al_{0.7}Ga_{0.3}As$ etch stop layer, and a second stopping at the subsequent GaAs layer. After etching off the back, a 700nm thick membrane remained, terminated with a GaAs layer and supported by the carrier wafer. To define the active region we etched a 500nm deep dumb-bell shaped mesa, 5μm wide at its narrowest point. A hole was etched through the remaining 200nm of AlGaAs to the center conductor of one of the cpw lines. (The 200 nm insulating layer is not functionally necessary but makes processing easier. It is not visible in the SEM micrographs of Fig. 1, and not shown in the schematic diagram of Fig. 1c) This hole was filled by subsequent metallization and electrically connected the top gate to the buried center conductor of one of the cpw lines. For measurement, the devices were diced into 2mm by 2mm chips and glued on a 12mm or 6mm diameter silicon lens.  The development of this fabrication process was the most time-consuming part of this project.

*Responsivity estimation*

In measuring the responsivity of device A, the peak voltage induced by FEL pulses was measured on an oscilloscope, averaged over the 5 μs pulse duration, and divided by the gain of the preamplifiers which had a 1 MHz bandwidth.  The power was measured by a calibrated energy meter placed at the position of the detector with all optics undisturbed but the cryostat removed. In order to calculate the power coupled to the antenna pattern, the Gaussian beam coupling between impinging radiation and equivalent Gaussian beam associated with the silicon lens has been considered. In particular, in our measurement set-up two different beam waists were considered: 0.5 +/- 0.25 mm for the impinging radiation and 4.89 mm for the silicon lens. Moreover, the effect of the distance between the beam waists and an axial misalignment between the two beams due to a small error in positioning our dewar were also considered. The simple formulation to account for these parameters can be found in reference [2].  The small size of the impinging beam waist compared with the beam waist for the Si lens is responsible for the poor estimated coupling efficiency quoted in the paper (4±2%).  Device A failed before the THz beam waist could be expanded to match that of the Si lens.

*Impedance matching*

A crucial design consideration is impedance-matching the quantum well absorbing region to the antenna for maximum power transfer.  On resonance, the quantum-well absorber dissipates the incident power like a resistor capacitively coupled to the coplanar waveguide and antenna. The total load impedance $Z_{Load}$ of the active region can be modeled as a resistor in series with the geometric capacitance of the active region. The resistance is given by[3]

$$R_{zz} = \frac{N_S e^2 f_{12} n(T)}{\varepsilon^2 \varepsilon_0^2 A m^* \omega^{*2} 2\Gamma}$$

where $N_s$ is the 2-D charge density, $e$ is the electronic charge, $f_{12}$ is the oscillator strength, $n(T) = (N_1 - N_0)/N_S$ is the normalized population difference between ground and excited subbands, $\varepsilon$ is the dielectric constant of GaAs, $\varepsilon_0$ is the permittivity of free space, A is the area of the active region, $m^* = 0.067 m_0$ is the effective mass of the conduction electron in GaAs, $\omega^*$ is the intersubband absorption frequency, and $\Gamma$ is the half-width at half-maximum of the intersubband absorption, as described in detail in Ref.[3]. The parameters $f_{12}$ and $n(T)$ were computed self-consistently in the Hartree approximation at 120 K and assuming $\Gamma$=0.25 THz. The high electron temperature relative to the intersubband spacing ($h$ x 1.53 THz/$k_B$=73K ) to makes $n(T)$ relatively small compared to its maximum value of 1.

The source impedance $Z_{Source}$=20$\Omega$-i40$\Omega$ is calculated taking into account the slot active impedances, the characteristic parameters of the CPW, radiation and conduction losses along the transmission lines and the RF choke filter impedances. The procedure to calculate all these contributions has been tested in the case of Hot Electron Bolometer mixers that have an RF circuit similar to the one used in the TACIT detectors[4].

*Methods to improve TACIT detectors*

Low-loss window materials, antireflection coatings, and Gaussian mode-matching can lower insertion losses. Future microwave designs can correct for the impedance mismatch between the antenna and the active region. For example, with the identical active region, the power transfer from the antenna to the load can be boosted from 4±0.4% to 45% by changing the antenna embedding circuit to have an inductive source impedance. This tunes out the capacitive load impedance of the active region at a chosen frequency. Together, these improvements can lower the Johnson-noise limited NEP for a device with an identical active region operating at a bath temperature of 80K to a minimum of $\eta\sqrt{4k_BT_eR}\big/\left(0.45\times R_{optical}\right)$=4x10$^{-11}$ W/Hz$^{1/2}$. Here we assume $R_{optical}/\eta$=107 V/W derived from optical measurements, $R$=1000 Ohms, $T_e$=120K.

The active region can also be redesigned. The critical parameter $\gamma$ can be enhanced by a factor of 3 by using a 2-DEG in a wide square well with low-temperature mobility in excess of $10^6$ cm$^2$/V-s (see inset to Fig. 4 and ref. [5]). Such a quantum well will have a narrower intersubband linewidth, and a larger $f_{12}$. A quantum well with higher mobility will also maintain its relatively large $\gamma$ down to relatively low electron temperatures, enabling operation at an electron temperature of the order of 60K rather than 120K in most of this work. Thus $n(T)$ will be increased. The enhancements to $f_{12}$ and $n(T)$ enable one to reduce the number of carriers in the active region and its area A while still matching the impedance of the antenna[3]. Using the expression for NEP found in Ref. 6 and assuming the self-consistent parameters A=10 μm$^2$, $T_e$=60K, $\tau$=10 ps, $\gamma$=0.03 and an overall coupling efficiency of 50%, one predicts NEP<10$^{-13}$ W/Hz$^{1/2}$. This supports the prediction at the end of the Letter of NEP ≤ 1pW/Hz$^{1/2}$.